\documentstyle[12pt]{article}
\title{Rare decays $B \to X_{s,d}\, \nu \bar \nu$ and 
$B_{s,d} \to l^+ l^-$ in the Topcolor-assisted Technicolor Model
\thanks{Project 19575015 is supported by the National Natural Science 
Foundation of China.} }
\author{ Xiao Zhenjun$^{1,2}$, 
Jia Liqun$^{3}$, L\"u Linxia$^{1}$ and Lu Gongru$^{1}$ \\  
{\small 1. Department of Physics, Henan Normal University,
Xinxiang, 453002 P.R.China.}\\ 
{\small 2. Department of Physics, Peking University, Beijing, 100871 
P.R.China.}\\ 
{\small 3. Department of Physics, Pingdingshan Teacher's College,}\\
{\small Pingdingshan, 467000 P.R.China.}\\ }
\date{\today}

\begin{document}
\maketitle
\begin{abstract}
We calculate the contributions to the rare decays $B \to X_{s,d}\, \nu 
\bar \nu$ and $B_{s,d} \to l^+l^- $ from one-loop $Z^0$-penguin diagrams
in the  framework of Topcolor-assisted Technicolor 
Model. Within the parameter space, we find that: 
(a) the new contribution from technipions is less than $2\%$ of 
the standard model prediction; (b) the top-pions 
can provide a factor of 10 to 30 enhancement to the ratios in question; 
(c) the topcolor-assisted technicolor model is consistent with the 
current experimental data.
\end{abstract}

\vspace{0.5cm}



%\end{document}
\newcommand{\beq}{\begin{eqnarray}}
\newcommand{\eeq}{\end{eqnarray}}

\newcommand{\paa}{P^{\pm}}
\newcommand{\pbb}{P_8^{\pm}}
\newcommand{\ptpm}{\tilde{\pi}^{\pm}}

\newcommand{\bsz}{b\overline{s}Z}
\newcommand{\bsd}{B \to X_{s,d} \nu \overline{\nu}}
\newcommand{\bsg}{B \to X_s \gamma }
\newcommand{\bs}{B \to X_{s} \nu \overline{\nu}}
\newcommand{\bd}{B \to X_{d} \nu \overline{\nu}}
\newcommand{\bbsd}{B_{s,d} \to l^+l^-}
\newcommand{\bbs}{B_{s} \to \mu^+  \mu^-}
\newcommand{\bbd}{B_{d} \to \mu^+ \mu^-}

\newcommand{\ssa}{\sin^2\theta_W}
\newcommand{\cca}{\cos^2\theta_W}
\newcommand{\mpa}{m_{p1}}
\newcommand{\mpb}{m_{p8}}
\newcommand{\mpi}{m_{\tilde{\pi}}}
\newcommand{\fpit}{F_{\tilde{ \pi}}}
\newcommand{\fpi}{F_{\tilde{ \pi}}}

\newcommand{\xzxt}{X_0(x_t)}
\newcommand{\cztc}{C_0^{New}}
\newcommand{\brbs}{B(B \to  X_{s}\nu \bar \nu)}
\newcommand{\brbd}{B(B \to X_{d} \nu \bar \nu)}
\newcommand{\brbsd}{B(B \to X_{s,d}\nu\bar\nu)}
\newcommand{\brbbs}{B(B_{s} \to \mu^+  \mu^-)}
\newcommand{\brbbd}{B(B_{d} \to \mu^+  \mu^-)}
\newcommand{\brbbsd}{B(B_{s,d} \to l^+  l^-)}

%\newpage

\vspace{1cm}
\section{ Introduction}

The examination of indirect effects of new physics in Flavor Changing
Neutral Current(FCNC) processes in B decays offers a 
complementary approach to the search for direct production of new particles 
at high energy colliders. Several FCNC transitions have been measured in B
system. Many more are accessible at the present and planned colliders(CESR, 
LEP/SLC, Tevatron, B factories, HERA-B/LHC)\cite{buras974}.

The new loop effects on the $B^0_d-\overline{B^0_d}$ mixing due to topcolor 
interactions has been estimated in refs.\cite{hill95,buchalla96}. 
And an upper bound can be placed on $\delta_{bd} \equiv |D_{Lbd}D_{Rbd}|$: 
$ \delta_{bd}/m_H^2 < 10^{-12} {\rm GeV}^{-2}$, which is an important constraint 
on the mixing factors. 

In the Standard Model(SM), the rare decays $\bsd$ and $\bbsd$
are theoretically very clean\cite{buras974}. 
The charm contribution is fully negligible, and the uncertainties 
related to the renormalization scale dependence can also be neglected. 
Consequently, these rare B-decay modes, as well as other clean rare K- 
and B-decays may play an important role in 
searching for the new physics beyond the SM.

In ref.\cite{xiao96}, the authors studied the new physics effects in the 
rare decays $\bsg$ and found that the Multiscale Walking Technicolor Model 
(MWTCM) \cite{lane91} was ruled out by the CLEO data\cite{cleo95} but the 
Topcolor-assisted Technicolor (TC2) Model\cite{hill95}  
is still consistent with the data. 
In refs.\cite{lu97}, the authors studied the new physics effects in the rare 
decays $B \to (K, K^*) l^+ l^-$, and found that the new physics effects 
may be measurable at future experiments. In ref.\cite{xiao98a} we found that
the MWTCM was ruled out by the rare K-decay data\cite{k98}. 
In ref.\cite{xiao98b}
 we calculated the contributions to rare decays $B \to X_{s,d}\, 
\nu \bar \nu$ and $B_{s,d} \to l^+l^- $ from charged technipions $\paa$ and 
$\pbb$ in the framework of One Generation Technicolor Model (OGTM) 
\cite{farhi} and MWTCM\cite{lane91}, 
respectively. We found that the first model is still consistent with the rare 
B-decay data\cite{aleph96,l397}, but the MWTCM is strongly disfavored 
by the data of $\brbs$\cite{aleph96}. 

In this paper, we will investigate the contributions to the rare B decays 
$B \to X_{s,d} \nu \bar \nu$ and $B_{s,d} \to l^+l^- $ from one-loop 
$Z^0$-penguin diagrams induced by the charged top-pions and technipions
appeared in the TC2 Model\cite{hill95}. 

This paper is organized as follows. In Sec.2 we  
extract out the new effective $Z^0$-penguin couplings. 
In Sec.3 and Sec.4, we present the numerical results for the 
branching ratios $\brbsd$ and $\brbbsd$ with the inclusion of new physics 
effects, respectively. The conclusions  are also 
included in the section 4.

%%%%%%%
\section{  TC2 models and new effective couplings}

Besides the Hill's TC2 model\cite{hill95}, other similar models with 
different fermion contents and  gauge group structures also proposed 
recently\cite{lane98}. But the basic ideas in all these models are the 
same: Firstly, the electroweak symmetry is  broken by technicolor with 
an extended 
technicolor (ETC), the large top quark mass is a combination of a dynamical 
condensate component $m_t^*=(1-\epsilon)m_t$, generated by the new strong 
topcolor interaction,
together with a small fundamental component, $m_{t1}=\epsilon m_t$ ($ 
\epsilon \ll 1$), generated by the ETC.   
Secondly, the existence of the top-pions is an essential feature in the 
Topcolor scenarios\cite{lane98}, regardless of the differences between 
models constructed so far. Finally, ordinary technipions should exist 
in all such  models.  

In TC2 model, the color-octet "coloron" $V_8$ 
(i.e., the top-gluons) and the color-singlet $Z'$ should be heavier than 
1 TeV\cite{burdman97,su97}. The three top-pions are nearly degenerate. If the 
top-pion is lighter than the top quark,  then  one has $\Gamma(t \to 
\tilde{\pi}^+ b) \approx (m_t^2 - \mpi^2)^2/(16\pi m_t \fpit^2)$. 
At $2-\sigma$ level, the lower bound is $\mpi \geq 100 {\rm GeV}$ from 
the Tevatron data\cite{pdf98}. The relatively light top-pions and other 
bound states, may provide potentially large loop effects in 
low  energy observables. This is the main 
motivation for us to investigate the contributions to the rare B-decays 
$\bsd$ and $\bbsd$ from the top-pions and technipions in the framework of 
TC2 model.

The couplings of the charged top-pions to t- and b-quarks take the form
\cite{hill95}: 
\beq
\frac{m_t^*}{ \fpit } \left[   
i \overline{t}_R b_L \tilde{\pi}^+
+ i \overline{b}_L t_R \tilde{\pi}^- \right],  \ \ \ 
\frac{m_b^*}{ \fpit } \left[   i\overline{t}_L b_R \tilde{\pi}^+
+ i \overline{b}_R t_L \tilde{\pi}^- \right]
\eeq
here, $m_t^*$ denotes the top quark mass generated by topcolor interactions,  
while $m_b^*$) is the bottom quark mass generated by $SU(3)_1$ instanton 
effects\cite{hill95}.

In TC2 models,
If one uses the square root of the CKM mixing matrix for $(U_L, D_L$)  
and assumes that the $U_R$ and $D_R$ are approximately diagonal, the 
constraints from the $B^0-\overline{B^0}$, $b \to s\gamma$ and 
$D^0-\overline{D^0}$ can  be avoided\cite{buchalla96}.  In this paper we 
use the square root of the CKM mixing matrix for $D_L$ and assume  that the 
$U_R$ and $D_R$ are simply diagonal, which means that the possible 
contributions from so-called $``b-pions"$ appeared in TC2 model must be 
very small and can be neglected safely. The mixings between the third and 
fist two generation quarks are therefore can be written as:
\beq
\frac{m_t^*}{\fpit} \left[ 
i\tilde{\pi}^+ \overline{t}_R (d_L D_{Ltd} +  s_L D_{Lts}) + h.c. \right].
\eeq

In TC2 models, The new contributions to the rare B decays from technipions 
are suppressed roughly by a factor of $(m_{t1}/F_\pi)^2/ 
(m^*_t/\fpit)^2 \sim 10^{-3}$, when compared with that from the top-pions. 

The relevant gauge couplings of charged technipions and top-pions to $Z^0$ 
gauge boson are basically model-independent and can be found for instance 
in ref.\cite{eichten86}.
The effective Yukawa couplings of charged technipions to
fermion pairs can be found in refs.\cite{xiao98a,ellis81,eichten86}.

The corresponding one-loop diagrams in the SM were evaluated long time 
ago and can be found in ref.\cite{inami81}. 
The new penguin diagrams can be obtained by replacing the internal $W^{\pm}$ 
lines with the unit-charged top-pion and technipion lines\cite{xiao98b}.
The color-octet $p_8^{\pm}$ does not couple to the $l\nu$ lepton pairs, 
and therefore does not present in the box diagrams.
For the color-singlet technipion and top-pion, they do couple 
to $l\nu$ pairs through box diagrams, but the 
relevant couplings are strongly suppressed  by the lightness of $m_l$.
Consequently, we can safely neglect  the tiny contributions from 
technipion and top-pion through the box diagrams.

Because of the lightness of the $s$, $d$ and $b$ quarks  when 
compared with the 
large top quark mass and the technipion masses we set $m_d=0$, $m_s=0$, 
$m_b^*=0$ and $m_b=0$ in the calculation. We will use dimensional 
regularization to regulate all the ultraviolet divergences in the 
virtual loop corrections and adopt the Modified Minimal Subtraction 
($\overline{MS}$) renormalization scheme.  It is easy to show that all 
ultraviolet divergences are canceled for $\paa$, $\pbb$ and $\ptpm$ 
respectively,  and therefore the total sum is finite.

By analytical evaluations of the Feynman diagrams, 
we find the effective $\bsz$ vertex induced by the charged top-pion   
exchange,
\beq
\Gamma^{I}_{Z_{\mu}} = 
\frac{1}{16 \pi^2}\frac{g^3}{\cos\theta_W}\; \sum_{j} \lambda_j\,
\overline{s_L}\, \gamma_{\mu}\, b_L\; C_0^{New}(\xi_j), 
\label{bsza}  \hspace{4cm}  \\
C_0^{New}(\xi_j) =\frac{D_{Ljs}^+ \mpi^2 \xi_j}{2\sqrt{2} 
     V_{js} \fpit^2 G_F M_W^2} \left[ \frac{(-1 +2\ssa -3\xi_j 
+2\ssa \xi_j)}{8(1-\xi_j)} - \frac{\cca\xi_j \ln[\xi_j]}{2(1-\xi_j)^2}
\right]
\label{cza}
\eeq
where $\lambda_j=V_{js}^*V_{jb}$, $\xi_t=m_t^{*2}/\mpi^2$, 
$\xi_j=m_j^2/\mpi^2$ for $j=c, u$, $\sin\theta$ is the Weinberg angle, 
$M_W$ is the W boson mass and $G_F$ is the Fermi coupling constant. 
For the case of the effective 
$b\bar d Z$ vertex, the $s$ in eqs(\ref{bsza},\ref{cza}) should be replaced 
by $d$. In TC2 models, one usually uses $\fpi = 50 \sim 70 {\rm GeV}$
\cite{hill95,buchalla96}.

For the case of technipions, the functions 
$C_0^{New}(y_j)$ and $C_0^{New}(z_j)$ are 
\beq
C_0^{New}(y_j)=\frac{\mpa^2 y_j}{3\sqrt{2}F_{\pi}^2 G_F M_W^2} 
\left[ \frac{(-1 +2\ssa -3y_j +2\ssa y_j)}{8(1-y_j)} - 
\frac{\cca y_j \ln[y_j]}{2(1-y_j)^2}\right]\label{czb}    \\
C_0^{New}(z_j)=\frac{8 \mpb^2 z_j}{3\sqrt{2}F_{\pi}^2 G_F M_W^2}
\left[ \frac{(-1 +2\ssa -3z_j +2\ssa z_j)}{8(1-z_j)} -
 \frac{\cca z_j \ln[z_j]}{2(1-z_j)^2}\right]
\label{czc} 
\eeq
where  $y_t=m_{t1}^2/\mpa^2$, $z_t=m_{t1}^2/\mpb^2$ and $F_{\pi}=123 
{\rm GeV}$ is the technipion weak decay constant.
In the above calculations, we used the unitary relation $\sum_{j=u,c,t}
\lambda_j\cdot \;constant=0$ wherever possible, and  neglected the masses 
for all external lines. We also used the functions $(B_0, B_{\mu}, 
C_0, C_{\mu}, C_{\mu\nu})$ whenever needed to make the integrations, and the
explicit forms of these complicated functions can be found, for instance, 
in ref.\cite{cho91}.

Within the standard model, the rare B-decays under consideration depend on 
the functions $X(x_t)$ and/or $Y(x_t)$ ($x_t=m_t^2/m_W^2$), 
they are currently known at the NLO level \cite{buras974}.
When the new contributions from charged technipions  and top-pions
are included, the functions $X$, and $Y$  can be written as
\beq
X &=& X(x_t) + C_0^{New}(\xi_t) + C_0^{New}(y_t) + C_0^{New}(z_t) ,
\label{xxtt} \\
Y &=& Y(x_t) + C_0^{New}(\xi_t) + C_0^{New}(y_t) + C_0^{New}(z_t). 
\label{yxtt}
\eeq

In the numerical calculations, we  fix the relevant parameters as follows 
and use them as the standard input(SIP)\cite{buras974,pdf98}: $M_W=80.41
{\rm GeV}$,  $G_F=1.16639\times 10^{-5} {\rm GeV}^{-2}$, $\alpha=1/129$,
$\ssa =0.23$, $m_t \equiv\overline{m_t}(m_t) = 170{\rm GeV}$, $\tau(B_{s})=
\tau(B_{d})=1.6ps$, $\Lambda^{(5)}_{\overline{MS}}=0.225{\rm GeV}$,
$F_{B_s}=0.210{\rm GeV}$, $m_{B_s}=5.38{\rm GeV}$, $m_{B_d}=5.28{\rm GeV}$,
$A=0.84, \;  \lambda=0.22, \;  \rho=0,\;  \eta=0.36$.
For $\alpha_s(\mu)$ we use the two-loop expression as given in 
ref.\cite{buras974}.

In the SM, using the SIP, we have $X(x_t)=1.54$, $Y(x_t)=1.06$.  
For $\mpa=(50\sim 250){\rm GeV}$,  $\mpb=(100\sim 600){\rm GeV}$ and 
$\epsilon =(0.05 \sim 0.1)$,  we have $|C_0^{New}(y_t)| \leq 5.5 \times 
10^{-4}$, $|C_0^{New}(z_t)| \leq 1.8 \times 10^{-2}$. 
For $\mpi=(100\sim 350){\rm GeV}$ and $\epsilon =0.05$ ($0.1$), we have  
$|C_0^{New}(\xi_t)| = 4.52\sim 0.81$ ($3.91 \sim 0.61$), which is much larger
than $|C_0^{New}(y_t)|$ and $|C_0^{New}(z_t)|$.  
Consequently, top-pions $\ptpm$ will dominate the new contribution. 

%%%%%%%%%%%%%%%%%%%%
\section{ The decay $\bsd$ }

Within the Standard Model, 
the effective Hamiltonian for $\bs$  are now available at the NLO level
\cite{buras974}, 
\beq
{\cal H}_{eff} =  \frac{G_F}{\sqrt{2}}\frac{\alpha}{2\pi \ssa} V^*_{tb}
V_{ts} X(x_{t}) (\overline{b}s)_{V-A}(\overline{l}l)_{V-A}+h.c.  
\label{heff}
\eeq
with {\it s} replaced by {\it d} in the case of $\bd$. 
 
Using the effective Hamiltonian $(\ref{heff})$, normalizing to 
$B(B \to X_{c} e \overline{\nu})$ and summing over the three 
neutrino flavors one finds
\beq
B(\bs) = B(B \to X_{c} e \overline{\nu}) \frac{3 \alpha^2}
{4 \pi^2\sin^4\theta_{W}} \frac{\mid V_{ts} \mid ^2}{\mid V_{cb} \mid ^2} 
\frac{X(x_{t})^2}{f(z)} \frac{\overline{\eta}}{\kappa(z)} 
\eeq
where $f(z)=0.542$, $\kappa(z)=0.880$ and $\overline{\eta}=0.831$ for
$z=m_c/m_b=0.29$. In the case of $\bd$ one has to replace $V_{ts}$
by $V_{td}$ which results in a decrease of the branching ratio by roughly an
order of magnitude.

Within the SM, using the SIP, and setting 
$B \to X_{c}e \overline{\nu}=10.5\%$ and $\mid V_{ts}/V_{cb} \mid ^2=0.95$,
one finds 
\beq
\brbs^{SM}=3.54 \times 10^{-5},  \ \ \ 
\brbd^{SM}=2.04 \times 10^{-6}
\eeq
which is consistent with the result given in ref.\cite{buras974}. 

Using the SIP,  and assuming $\fpi=50{\rm GeV}$, $\epsilon=0.05$ 
and $100{\rm GeV} \leq \mpi \leq 350{\rm GeV}$, one finds
\beq
8.19 \times 10^{-5} \leq \brbs \leq 5.48 \times 10^{-4},  \\
3.2 \times 10^{-6} \leq \brbd \leq 1.24 \times 10^{-5}.
\eeq

In Fig.1, the solid (short-dash) curve shows the theoretical 
prediction when new contributions from technipions and top-pions 
are all included for $\epsilon=0.1$ ($0.05$). 
The upper dots line corresponds to the ALEPH 
data\cite{aleph96}: $ \brbs < 7.7 \times 10^{-4}$,  
which is a factor of 20 above the SM expectation (dot-dash line),
and is still consistent with the theoretical expectations when the 
new contributions from  the charged technipions and top-pions are included. 
For larger $\fpi$, the size of the new contribution from top-pions will be 
decreased accordingly.
For the decay $\bd$ no experimental bound is available currently.

\section{ The decay $\bbsd$ }

The effective Hamiltonian for $B_{s} \to l^+l^-$ is known at the NLO level 
\cite{buras974},
\beq
{\cal H}_{eff}=- \frac{G_{F}}{\sqrt{2}}  \frac{\alpha}{2 \pi \ssa} 
V^{*}_{tb} V_{ts} Y(x_{t}) (\overline{b} s)_{V-A} (\overline{l}l)_{V-A}+h.c.
\label{hef}
\eeq
with $s$ replaced by $ d$ in the case of $B_d \to l^+l^-$.

Using the effective Hamiltonian (\ref{hef}) and summing over three neutrino
flavors one finds
\begin{eqnarray}
\brbbs=\tau(B_{s}) \frac{G_{F}^2}{\pi} (\frac{\alpha}{4 \pi \ssa})^2
F_{B_{s}}^2 m^2_{l} m_{B_{s}} \sqrt{1-4 \frac{m_{l}^2}{m^2_{B_{s}}}}
\mid V_{tb}^{*} V_{ts} \mid^2 Y(x_{t})^2
\end{eqnarray}
with $s$ replaced by $ d$ in the case of $B_d \to l^+l^-$, and $B_s$ ($B_d$) 
denotes the flavor eigenstate $\overline{b}s$ $(\bar b d)$ and $F_{B_s}$ 
$(F_{B_d})$ is the corresponding decay constant.

Again the new contributions from technipions are very small. The dominant 
enhancement due to the unit-charged top-pion 
can be as large as a factor of 30 at the level of the SM predictions.
The numerical results for various decay modes are given in Table 1. 
In the numerical calculations, we use the SIP, 
assuming $100{\rm GeV} \leq \mpi \leq 350{\rm GeV}$,$\mpa=50{\rm GeV}$, 
 $\mpb=100{\rm GeV}$,  and setting 
$\mid V^{*}_{tb}V_{ts} \mid ^2=0.0021$, $\mid V^{*}_{tb}V_{td} \mid ^2
=1.3 \times 10^{-4}$, and $D_{Lts}^+/V_{ts}=D_{Ltd}^+/V_{td}=1/2$.

The currently available experimental bounds are\cite{l397}:  
$B(B_s \to e^+ e^-) < 5.4 \times 10^{-5}$ at $90\% C.L.$, 
$B(B_s \to \mu^+ \mu^-) < 2.0 \times 10^{-6}$ at $90 \% C.L.$,  
and $\brbbd < 6.8 \times 10^{-7}$ at the $90\% C.L$. 
It is easy to see that the current experimental bounds are still about 2 
orders of magnitude away from
the theoretical prediction  even if the large enhancements due
to charged top-pions are taken into account. CDF and BarBar may reach 
the sensitivity of $1 \times 10^{-8}$ and $4 \times 10^{-8}$ for 
$\brbbs$ and $\brbbd$ in the near future. Such sensitivity is on the margin 
to probe the effects due to unit-charged top-pions.

\begin{table}
\caption{The branching ratios of $\bbsd$, and 
$\epsilon=0.05$. }
\begin{center}
\begin{tabular}{|c|c|c|c|c|}
\hline
{\rm Branching ratio} & SM & plus technipion & SM + New & Data \\
\hline
{$B(B_{s} \to e^+ e^-)$} & $0.72\times 10^{-13}$ & $0.71\times 10^{-13}$ 
& $(0.22 \sim 2.01 ) \times 10^{-12}$ & $ < 5.4 \times 10^{-5}$ \\ \hline
{$B(B_{s} \to \mu^+ \mu^-) $} & $3.09\times 10^{-9}$ 
& $3.05\times10^{-9}$ & $(0.96 \sim 8.58)\times 10^{-8}$ & $
< 2.0 \times 10^{-6}$\\ \hline
{$B(B_{s} \to \tau^+ \tau^-)$} & $0.66\times 10^{-6}$ & $0.65 \times 10^{-6}$ 
& $(0.20 \sim 1.82)\times 10^{-5}$ &  \\
\hline 
{$B(B_{d} \to e^+e^-) $} & $0.44\times 10^{-14}$ & $0.43 \times 10^{-14}$ 
& $(0.16 \sim 1.22) \times 10^{-13}$&  \\
\hline 
$B(B_{d} \to \mu^+ \mu^-) $ & $1.88\times 10^{-10}$ & $1.85\times 10^{-10}$ 
& $(0.58 \sim 5.21) \times 10^{-9}$ & $< 6.8\times 10^{-7}$ \\
\hline
$B(B_{d} \to \tau^+ \tau^-) $ & $0.40\times 10^{-7}$ & 
$0.39\times 10^{-7}$ & $(0.12 \sim 1.11) \times 10^{-6}$ & \\ 
\hline
\end{tabular}
\end{center}
\end{table}

In summary, the contributions due to technipions are less than $2\%$ of the 
SM predictions and can be neglected safely. Within the considered parameter 
space, the top-pions can provide a factor of 10 to 30 enhancement to the 
branching ratios in question. The theoretical prediction of $\brbs$ is now 
close to the experimental bound as illustrated in Fig.1. The TC2 model is 
still consistent with currently available data. Further improvement in the 
sensitivity of the relevant data will be very helpful to find the signal 
of charged top-pions or put some limits on their mass spectrum.

%\newpage

%\newpage

\vspace{1cm}
\begin{center}
{\bf Figure Captions}
\end{center}
\begin{description}

\item[Fig.1:] Plots of the branching ratios $\brbs$ vs the mass $\mpi$.
For more details see the text.

\end{description}


\vspace{1cm}
\begin{thebibliography}{99}

\bibitem{buras974}
A. J. Buras and R.Fleischer, in {\it Heavy Flavor II}, 
edited by  A.J.Buras and M.Lindner (World Scientific Publishing Co., 
Singapore, 1998);
G.Buchalla, A.J.Buras and M.E.Lautenbacher,  Rev.Mod.Phys. {\bf 68}
(1996)1125.

\bibitem{hill95}
C. T. Hill, Phys.Lett. {\bf B 345} (1995)483; 
K. Lane and E. Eichten, Phys.Lett. {\bf B 352}(1995)382;
D. Kominis, Phys. Lett. {\bf B 358}(1995)312;

\bibitem{buchalla96}
G.Buchalla, G. Burdman, C.T.Hill and D. Kominis, Phys.Rev. 
{\bf D53}(1996)5185.

\bibitem{xiao96}
C.D. L\"u, Z.J. Xiao, Phys.Rev. {\bf D53}(1996)2359; 
G.R. Lu, Y.G. Cao, Z.H. Xiong and C.Y. Yue, Z.Phys. C{\bf 74}(1997)355;
G.R. Lu, Y.G. Cao, Z.H. Xiong and Z.J.Xiao, 
{\it High Energy Phys. and Nucl. Phys. (in Chinese)}, 21(1997)1005; 
G.R. Lu, J.S. Huang, Z.J. Xiao and C.X. Yue, Commun.Theor.Phys. to be 
published. 


\bibitem{lane91}
K.Lane and M.V. Ramana, Phys.Rev. {\bf D44}(1991)2678. 

\bibitem{cleo95}
CLEO Collaboration, M.S.Alam {\it et al.}, Phys.Rev.Lett. {\bf 74}(1995)2885. 

\bibitem{lu97}
G.R. Lu, Z.H. Xiong and Y.G. Cao, Nucl.Phys. {\bf B487}(1997)43;
C.Greub, A.Ioannissian and D. Wyler, Phys.lett. {\bf B346}(1995)149.

\bibitem{xiao98a}
Z.J. Xiao, L.X.L\"u, H.K. Guo and G.R. Lu, Eur.Phys.J. {\bf C7}(1999)487.

\bibitem{k98}
S. Adler et al., Phys. Rev. Lett. {\bf 79}(1997)2204; 
E791 Collaboration, A.P.Heinson {\it et al.}, Phys. Rev. {\bf D51}(1995)985.

\bibitem{xiao98b}
Z.J. Xiao, J.Y. Zhang, L.X. L\"u and G.R. Lu, {\it High Energy Phys. and 
Nucl. Phys. (in Chinese)}, Vol.23(1999),  in press; 
Z.J. Xiao, L.X. Lu, H.K. Guo and G.R. Lu, Chin.Phys.Lett. Vol.16(1999)88.


\bibitem{farhi}
E. Farhi and L.Susskind, Phys. Rev. {\bf D20}(1979)3404;
S.Dimopoulos and L. Susskind, Nucl. Phys. {\bf B155}, 237(1979). 

\bibitem{aleph96}
ALEPH Collaboration, H.Kroha et al., Proceedings of the 28th 
International Conference on High Energy Physics, edited by Z.Ajduk 
and A.K.Wroblewski, (World Scientific Publishing Co., Singapore, 
1997), p.1182.

\bibitem{l397}
L3 Collaboration, M. Acciarrit {\it et al.}, Phys.Lett. {\bf B391}(1997)474; 
CDF Collaboration, F.Abe {\it et al.}, Phys.Rev. {\bf D57}(1998)R3811. 


\bibitem{lane98}
K.Lane Phys.Lett. {\bf B433}(1998)96, and references therein.

\bibitem{burdman97}
G. Burdman, D. Kominis, Phys. Lett. {\bf B403}(1997)101; 

\bibitem{su97}
Y. Su, F. Bonini and D.Kominis, Phys. Rev. Lett. {\bf 79}(1997)4075.

\bibitem{2hdm}
S.L.Glashow and S.Weinberg, Phys.Rev. {\bf D15}(1977)1958.

\bibitem{eichten86}
E.Eichten, I.Hinchliffe, K.Lane and C.Quigg, Rev.Mod.Phys. {\bf 56}(1984)579;
Phys.Rev. {\bf D34}(1986)1547; T.P.Cheng and L.F. Li, {\it 
Gauge theory of elementary particle physics}, Clarendon 
Press, Oxford, 1984.

\bibitem{ellis81}
J.Ellis,M.K.Gaillard, D.V.Nanopoulos and P.Sikivie, Nucl.Phys. {\bf B182}
(1981)505.

\bibitem{inami81}
T.Inami and C.S. Lim, Prog.Theor.Phys. {\bf 65}(1981)297.

\bibitem{cho91}
P.Cho and B.Greistein, Nucl.Phys. {\bf B365}(1991)279.

\bibitem{pdf98}
Particle Data Group, C. Caso {\it et al.}, Eur. Phys. J. {\bf C3}(1998)1.

\end{thebibliography}
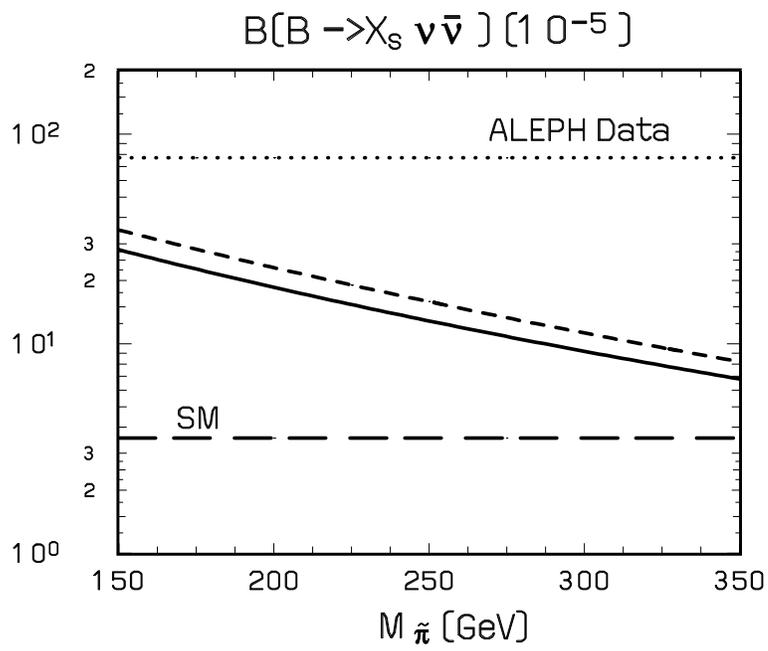
\end{document}